\documentclass[12pt]{ijck}

\usepackage[english]{babel}
\usepackage{csquotes}
\usepackage{textcomp}
\usepackage{latexsym,amsmath,amssymb}

\usepackage{fancyvrb,newverbs,xcolor}
\usepackage{graphicx}
\usepackage{caption,subcaption}

\definecolor{cverbbg}{gray}{0.93}
\colorlet{pyyellow}{yellow!10!white}
\colorlet{yamlblack}{black!10!white}

\newverbcommand{\yabox}
  {\setbox\verbbox\hbox\bgroup}
  {\egroup\colorbox{yamlblack}{\box\verbbox}}
\newverbcommand{\pybox}
  {\setbox\verbbox\hbox\bgroup}
  {\egroup\colorbox{pyyellow}{\box\verbbox}}

\usepackage[version=4]{mhchem}

\usepackage{siunitx}
\DeclareSIUnit\atm{atm}

\newcommand\ck{ChemKED}
\newcommand\pk{PyKED}

\usepackage[inline]{enumitem}
\setlist{noitemsep}

\usepackage{microtype}
\title{ ChemKED: a human- and machine-readable data standard for chemical kinetics experiments }

\author[1]{Bryan W.~Weber}
\author[2]{Kyle E.~Niemeyer\thanks{Corresponding author: \email{kyle.niemeyer@oregonstate.edu}}}

\affil[1]{Department of Mechanical Engineering, University of Connecticut, Storrs, CT, USA}
\affil[2]{School of Mechanical, Industrial, and Manufacturing Engineering, Oregon State University, Corvallis, OR, USA}

\begin{document}
\maketitle


%====================================================================
\begin{abstract} % not to exceed 200 words
    Fundamental experimental measurements of quantities such as ignition delay times, laminar flame
    speeds, and species profiles (among others) serve important roles in understanding fuel
    chemistry and validating chemical kinetic models. However, despite both the importance and
    abundance of such information in the literature, the community lacks a widely adopted standard
    format for this data. This impedes both sharing and wide use by the community. Here we introduce
    a new chemical kinetics experimental data format, ChemKED, and the related Python-based package
    for validating and working with ChemKED-formatted files called PyKED. We also review past and
    related efforts, and motivate the need for a new solution. ChemKED currently supports the
    representation of autoignition delay time measurements from shock tubes and rapid compression
    machines. ChemKED-formatted files contain all of the information needed to simulate experimental
    data points, including the uncertainty of the data. ChemKED is based on the YAML data
    serialization language, and is intended as a human- and machine-readable standard for easy
    creation and automated use. Development of ChemKED and PyKED occurs openly on GitHub under the
    BSD 3-clause license, and contributions from the community are welcome. Plans for future
    development include support for experimental data from laminar flame, jet stirred reactor, and
    speciation measurements.
\end{abstract}

%====================================================================
\section{Introduction}
Fundamental combustion experiments provide vital data for understanding fuel
chemistry and validating chemical kinetic models. Important measured quantities
include autoignition delay times, laminar flame speeds, and species profiles,
among others. However, despite both the importance and abundance of such
information in the literature, the combustion\slash chemical kinetics community
lacks an accepted, commonly used standard for recording and sharing data from
fundamental combustion experiments.\footnote{Interestingly, this contrasts the
situation for chemical kinetic models, where the CHEMKIN~\autocite{Kee:1996ck}
format dominates. Competing standards such as the
Cantera~\autocite{Cantera:2.3.0} CTI format or
FlameMaster~\autocite{FlameMaster:ref} lag behind considerably,
although use of the former continually grows due to Cantera's open availability.
% Similarly, % with more space, discuss JANAF/NASA polynomial standard ?
}

Instead, such data is typically found in under-documented comma-separated value
(CSV) files and Excel spreadsheets, or contained in PDF tables rather than as
supplementary material associated with a paper. In the worst case, data is only
available in figures and must be digitized by the use of software such as
WebPlotDigitizer~\autocite{WebPlotDigitizer}. These practices limit wide use
of the valuable, extensive experimental data available in the literature.
\textcite{Frenklach:2007bm} further explained the benefits of standardized,
widely available data for combustion. In brief, much of fundamental
combustion\slash chemical kinetics research drives towards the ultimate goal of
developing predictive kinetic models---and ultimately this depends on a
\emph{community} infrastructure for data and methods.

The combustion community has not yet widely adopted a standard format for
experimental data, but some researchers and groups have proposed solutions to
this problem. Most notably, Frenklach developed the PrIMe (\textbf{Pr}ocess
\textbf{I}nformatics \textbf{M}od\textbf{e}l) data
format~\autocite{Frenklach:2007bm,You:2011hy} and an associated online database,
the PrIMe Data Warehouse (available at \url{http://primekinetics.org}).
PrIMe files encode fundamental
combustion experimental data, along with kinetic models and calculated
quantities, using the eXtensible Markup Language (XML) standard.

PrIMe~\autocite{Frenklach:2007bm,You:2011hy}
has a number of features that make
it a strong standardization format, but it suffers from several flaws that
prevent wide adoption. First, the PrIMe standard does not require or support all
the information needed to simulate an experiment, including a machine-readable
definition of ignition or a standard way to express detailed facility-specific
effects necessary to properly simulate certain experiments. Second, PrIMe uses
internal identifiers for bibliographic references and
species\slash reactions, rather than standard identifiers (e.g., DOIs for
scholarly products, InCHI or SMILES for species). This limits the usability of PrIMe files outside
the PrIMe ecosystem.

Third, XML is intended to be a machine-readable markup language rather than a data format.
As a general-purpose document markup language, XML can represent detailed
information about a dataset, but implementing this detail requires XML-formatted
files to be much more verbose than files formatted with a language focused
only on representing data structures. The verbosity inherent in the XML format
limits the human readability of database files constructed in XML and presents a barrier to
creating and working with them.

Finally, the closed and opaque natures of the PrIMe standard and associated Data Warehouse limit
contributions and community development of the PrIMe standard and compatible data files and tools.
In this sense, we mean ``closed'' as a contrast to open source, open science, and open data, as
defined by the Open Knowledge Institute~\cite{openknowledge} or the Open Source
Initiative~\cite{opensource}. Although most commonly applied to software, the distinction between
open and closed is becoming increasingly important in science and research. In brief, the Open
Knowledge Institute defines open knowledge as ``free to access, use, modify, and
share\textellipsis — subject, at most, to measures that preserve provenance and
openness''~\cite{openknowledge}, while the Open Source Institute focuses on software and declares
that a project must provide free and unrestricted access to the source code, allow free
redistribution, permit derived works, and not discriminate
against any people or areas of work, along with a few other criteria~\cite{opensource}.

More recently, Varga and coworkers~\autocite{Varga2015a,Varga2017} developed the
ReSpecTh standard (available at
\url{http://respecth.chem.elte.hu/respecth/reac/CombustionData.php}),
which builds on the PrIMe format. ReSpecTh adds important
features that make files better standalone representations of experimental
data---i.e., more informative by themselves rather than in concert with a larger
system. For example, where PrIMe uses internal bibliographic references,
ReSpecTh adds a field for typical bibliographic data, including the DOI of
the cited article. The ReSpecTh standard also
provides machine-readable formats for describing ignition experiments, including
a field for the definition of autoignition and the ability to specify
facility-specific effects. It can also describe speciation measurements in
perfectly stirred reactors, flow reactors, and burner-stabilized flames, as well as
burning velocity data. To support this, ReSpecTh files allow the specification of species by
standard identifiers, including InChI or SMILES. Finally, each ReSpecTh file
has a unique DOI and URL assigned to it.

However, ReSpecTh experimental files are another
XML-based format, and, as such, suffer from the same readability issues as PrIMe.
Moreover, potential users must register using a valid institutional email address
to access the growing database.

Nonetheless, databases of fundamental combustion experiments
have proved quite useful. For instance, Olm and colleagues used their
ReSpecTh-based database to quantify the performance of literature
hydrogen~\autocite{Olm:2014gn} and syngas~\autocite{Olm:2015ch} kinetic models,
and more recently develop improved models for hydrogen~\autocite{Varga:2015hy},
hydrogen\slash syngas~\autocite{Varga:2016gj},
methanol\slash formaldehyde~\autocite{Olm:2017a},
and ethanol~\autocite{Olm:2016et}. In all six cases, they converted numerous
experimental datasets from the literature into the ReSpecTh standard.
Their database currently contains \num{93326}
experimental data points in 1396 ReSpecTh XML files~\autocite{respecth}.

Despite the development of these standards, limited examples exist in the
literature of open data \emph{sharing} using the formats. To the best
of our knowledge, the largest such publicly available database is hosted by
the Clean Combustion Research Center at King Abdullah University of Science and
Technology (KAUST): CloudFlame (available at \url{https://cloudflame.kaust.edu.sa}).
Started in 2013~\autocite{Goteng:2013cf,Goteng:2014,ReynoChiasson:2015}, CloudFlame
serves as an openly accessible database for experimental data, available in a standard
CSV format, and also provides a cloud infrastructure for running simulations
based on stored models and data. While an admirable and useful effort,
CloudFlame, like ReSpecTh and PrIMe, is a closed system controlled by a single
institution rather than by the community at large.

We therefore believe there remains a need for an open, community-focused and
developed, combustion\slash chemical kinetics data format. In this work, we
present a new, open-source, human- and machine-readable data standard for
fundamental combustion experiments, \ck{}, and offer tools for easily working
with data encoded in this format. Niemeyer recently introduced an initial
version of the \ck{} format~\autocite{Niemeyer:2016wf}, and in this work we
further formalize and develop the standard. We also discuss a related Python
software tool for validating and working with \ck{} files: \pk{}. Our
motivations resemble those of the PrIMe and ReSpecTh teams: the \ck{} project
will enable easy sharing and use of fundamental combustion data, for the
(primary) purposes of developing and validating predictive chemical kinetic
models. However, we make usability a major design focus, and plan to share all
data and software openly to cultivate a community of user-contributors.

%====================================================================
\section{Overview of ChemKED format}
\label{sec:overview-of-format}

This section provides an introduction to the current version of the \ck{} database format, v0.3.0,
which represents a functional standard for documenting autoignition measurements in shock tubes and
rapid compression machines. Here we describe the fields that a user creating a \ck{} file is likely
to use. There are several fields that are used internally in the schema that are not described here,
but can be found in the online documentation for the
schema.\footnote{\url{https://pr-omethe-us.github.io/PyKED/schema-docs.html}} In
Section~\ref{sec:usage-example}, we provide examples of \ck{} files and representative use cases. We
indicate \yabox|yaml| keywords or values using text with gray background, and \pybox|Python| code
using text with a yellow background.

\ck{} files use the YAML data serialization format~\autocite{yaml:1.2}. This
format offers the advantages of being relatively easy for humans to read and write,
encoded in plain text (ASCII or UTF-8 format), and
having parsers in most common programming languages, including Python, C++, Java,
Perl, MATLAB, and many others. The YAML syntax is quite simple: the basic file
structure consists of mappings, delimited by a colon. The key for the mapping is
on the left of the colon, and the value on the right can be a single value,
a sequence of values, a nested mapping, or some combination of these:
\begin{yamlbox}
key1: value  # Single-value mapping
key2:  # Sequence format
  - value
  - value
key3:  # Nested mapping
  key4: 0
key5:  # Sequence of mappings
  - key-s1: 0
    key-s2: value
  - key-s2: value
    key-s1: 0
\end{yamlbox}

The \yabox|value| can be a string, integer, or floating-point number. We designed
the \ck{} format to include all information necessary to simulate a
given experiment. A \ck{} file is generally broken into two main sections: a section containing the
``meta'' information about the experiment and the \ck{} file itself, and a section containing the
actual experimental data to be encoded.

\subsection{Apparatus information and metadata}
\label{sec:apparatus-information-and-metadata}

We will first describe the fields in the ``meta'' section.
This section uses a mapping that is called the \yabox|author| mapping in the schema. As this mapping
is used for any fields that specify an author throughout the schema, we describe it first.
The \yabox|author| mapping contains the following fields:
\begin{itemize}
    \item \yabox|name| (required, string): The author's full name
    \item \yabox|ORCID| (optional, string): the author's unique ORCID code
\end{itemize}
In general, the meta information describes the experimental facility and type of experiment, the
reference from which the data was taken, and the author of the \ck{} file itself. For each of the
keys below, we have indicated whether the key is required or optional, and the type of value
associated with the key. The keys in the meta section include:
\begin{itemize}
    \item \yabox|apparatus| (required, mapping): Information about the specific experimental
    apparatus used to conduct the experiment. The fields in this mapping are:
    \begin{itemize}
        \item \yabox|facility| (optional, string): A unique name or identifier for the apparatus,
        if the institution has several that are similar
        \item \yabox|institution| (optional, string): The institution where the experimental
        apparatus was located when the experiments were performed
        \item \yabox|kind| (required, string): The type of apparatus used to
        perform the experiment. Currently, only \yabox|shock tube| or
        \yabox|rapid compression machine| are supported.
    \end{itemize}
    \item \yabox|chemked-version| (required, string): The version of the \ck{}
    schema to which this file conforms
    \item \yabox|experiment-type| (required, string): Currently, only
    \yabox|ignition delay| is supported
    \item \yabox|file-author| (required, \yabox|author|-type mapping): The
    author of the \ck{} file
    \item \yabox|file-version| (required, integer): The version of the \ck{}
    file
    \item \yabox|reference| (required, mapping): The reference information for
    the published article associated with the data in the file. The fields in this
    mapping are:
    \begin{itemize}
        \item \yabox|authors| (required, sequence): A sequence of \yabox|author|
        mappings
        \item \yabox|detail| (optional, string): A description of from where the
        data originated (e.g., figure or table number)
        \item \yabox|doi| (optional, string): The article DOI
        \item \yabox|journal| (optional, string): The name of the publishing
        journal
        \item \yabox|pages| (optional, string): The article pages
        \item \yabox|volume| (optional, integer): The journal volume number
        \item \yabox|year| (required, integer): The year of publication
    \end{itemize}
\end{itemize}

Several of these keys bear further discussion. The \yabox|chemked-version| is present in the file so
that PyKED can determine whether the file conforms to a supported version of the schema. Although
quite similar in name, the \yabox|file-version| represents the number of versions that have been
created of the current file; it is an integer that should be incremented whenever changes are made
to the file. Finally, the \yabox|reference| mapping is required in every file, even those that
represent data that has not been published. However, the \yabox|reference| mapping only requires the
\yabox|authors| sequence and the \yabox|year|, so that \ck{} files can be used internally within
research groups to represent unpublished data prior to publication.

\subsection{Ignition delay experimental data}
\label{sec:ignition-delay-experimental-data}

The second section of the file encodes the experimental data as a sequence of mappings. The
top-level key for the sequence is called \yabox|datapoints|. For the current version of the \ck{}
schema, only autoignition delay experiments are supported. The following information
is required in each element of the sequence:
\begin{itemize}
    \item \yabox|temperature| (required, sequence): The temperature of the
    experiment, with units and optionally uncertainty
    \item \yabox|ignition-delay| (required, sequence): The ignition delay of the
    experiment, with units and optionally uncertainty
    \item \yabox|pressure| (required, sequence): The pressure of the experiment,
    with units and optionally uncertainty
    \item \yabox|first-stage-ignition-delay| (optional, sequence): If two stages of ignition are
    present in the experiment, this is the length of the first stage of ignition, and the
    \yabox|ignition-delay| is then the overall ignition delay
    \item \yabox|equivalence-ratio| (optional, float): The value of the equivalence ratio
    \item \yabox|composition| (required, mapping): The composition of the
    mixture in the experiment, described via a nested mapping with the fields:
    \begin{itemize}
        \item \yabox|kind| (required, string): one of \yabox|mole fraction|, \yabox|mole percent|, or \yabox|mass fraction|
        \item \yabox|species|: sequence of mappings, with the fields:
        \begin{itemize}
            \item \yabox|species-name| (required, string): The name of the species
            \item \yabox|amount| (required, sequence): The mole fraction or percent, or the mass fraction, optionally with uncertainty
            \item \yabox|InChI|, \yabox|SMILES|, or \yabox|atomic-composition|
            (required, string or sequence): The InChI
            or SMILES string representing the molecule, or its atomic
            composition as a sequence
        \end{itemize}
    \end{itemize}
    \item \yabox|ignition-type| (required, mapping): The method used to detect
    ignition in the experiments. The required fields are
    \begin{itemize}
        \item \yabox|type| (required, string): How ignition delay was
        measured; one of \yabox|d/dt max| (indicates the ignition point was
        found at the maximum of the time derivative of the \yabox|target|),
        \yabox|max| or \yabox|min| (indicates the ignition point was at the
        maximum or minimum of the \yabox|target|), \yabox|1/2 max| (
        indicates the half-maximum point of \yabox|target|), or \yabox|d/dt max extrapolated|
        (indicates the maximum slope of the target extrapolated to the baseline)
        \item \yabox|target| (required, string): The target for the ignition
        \yabox|type| measurement; one of \yabox|temperature|, \yabox|pressure|,
        \yabox|OH|, \yabox|OH*|, \yabox|CH|, or \yabox|CH*|
    \end{itemize}
\end{itemize}

Each of the
quantities in an element must be specified with units. This should be done as a
single string associated with the first element of the sequence. The units of
the quantity are validated to ensure appropriate dimensionality
for the quantity. In addition, each of the quantities in an element can
optionally be assigned an uncertainty. This uncertainty can be either absolute
or relative, and is specified as an element of the sequence of the associated key.
For example, the absolute uncertainty of the temperature and the relative
uncertainty of the ignition delay might be specified as:
\begin{yamlbox}
datapoints:
  - temperature:
      - 1100 kelvin
      - uncertainty-type: absolute
        uncertainty: 10 kelvin
    ignition-delay:
      - 10 us
      - uncertainty-type: relative
        uncertainty: 0.1
    ...
\end{yamlbox}

Frequently, experimental series hold certain properties constant, such as initial mixture
composition or pressure, and use the same method to detect ignition delay. \ck{} files support a
\yabox|common-properties| convenience section where these common details may be defined once, and
referenced in each element of the \yabox|datapoints| sequence. This block uses the ability of YAML
files to define an anchor with the \yabox|&| symbol and refer to that section later with the
\yabox|*| symbol:
\begin{yamlbox}
common-properties:
  composition: &comp
    kind: mole fraction
    species:
      ...
  ignition-type: &ign
    ...
datapoints:
  - composition: *comp
    ignition-type: *ign
    temperature:
      - 1000 kelvin
  ...
\end{yamlbox}
The \yabox|common-properties| section is not required, but can save space and
help avoid errors when several data points share some common values. However, even if
properties are specified in the \yabox|common-properties| section, they must also be
explicitly specified and referenced in the \yabox|datapoints| section for each data point,
as seen in the previous code.

\subsubsection{Shock tube experiments}
\label{sec:shock-tube-experiments}

There is one additional key that can be used to describe the facility-specific effects during a
shock tube experiment:
\begin{itemize}
    \item \yabox|pressure-rise| (optional, sequence): The pressure rise in the driven section after
    the passage of the reflected shock, in dimensions of inverse time. Must include units and,
    optionally, uncertainty
\end{itemize}

\subsubsection{Rapid compression machine experiments}
\label{sec:rcm-experiments}

Rapid compression machine experiments typically report several additional pieces of information in
addition to the ignition delay. As for shock tube experiments, these describe some of the
facility-specific effects during an experiment. The keys to describe this information are:
\begin{itemize}
    \item \yabox|volume-history| (optional, mapping): Specify the volume history of the reaction
    chamber as a function of time during a rapid compression machine experiment. The fields in this
    mapping are:
    \begin{itemize}
        \item \yabox|volume| (required, mapping): A mapping describing the volume in the history\footnote{Future versions
        of \ck{} will support specifying additional time histories for rapid compression machine and shock tube
        experiments (e.g., pressure, volume, light emission, OH emission). These will follow the specification format given for volume.}
        The fields in this mapping are:
        \begin{itemize}
            \item \yabox|units| (required, string): the units of the volume, with dimensions length
            cubed
            \item \yabox|column| (required, integer): the zero-based index of the column storing the
            volume information in the \yabox|values| array (e.g., 0 or 1)
        \end{itemize}
        \item \yabox|time| (required, mapping): A mapping describing the time in the history.
        The fields in this mapping are:
        \begin{itemize}
            \item \yabox|units| (required, string): the units of the time
            \item \yabox|column| (required, integer): the zero-based index of the column storing the
            time information in the \yabox|values| array (e.g., 0 or 1)
        \end{itemize}
        \item \yabox|values| (required, sequence): A sequence of time-volume pairs describing the values
        of the volume at different times
    \end{itemize}
    \item \yabox|compressed-temperature| (optional, sequence): The estimated temperature at the end
    of the compression stroke
    \item \yabox|compression-time| (optional, sequence): The time taken during the compression
    stroke
    \item \yabox|compressed-pressure| (optional, sequence): The measured pressure at the end of the
    compression stroke
\end{itemize}

\subsection{Public database of \ck{} files}
\label{sec:public-database}

We have begun building an open repository of \ck{} files, available online at
GitHub.\footnote{\url{https://github.com/pr-omethe-us/ChemKED-database}}
We welcome submissions from the community; these can be made via the standard pull request
on GitHub. The authors also encourage questions and problems to be submitted either via
email or as issues on GitHub. In particular, researchers interested in converters for
internal data formats to the \ck{} format are encouraged to reach out;
the \pk{} package (described in more detail in Sec.~\ref{sec:pyked-architecture})
currently provides converters to and from the ReSpecTh format~\autocite{Varga2017}.
While we plan to continue growing this database---and hope for submissions from
the community---others can freely download files, or in fact copy (``fork'' in GitHub
parlance) the entire database for their own purposes.

%====================================================================
\section{PyKED architecture}
\label{sec:pyked-architecture}
\pk{} is a Python package that provides the reference implementation
of the interface to \ck{} files~\autocite{PyKED}. \pk{} reads \ck{} files, validates
their structure and content, and allows the user to interact with the data contained
in the \ck{} file.

\pk{} provides the basic user interface to a \ck{} file through the
\pybox|ChemKED| class. The \pybox|ChemKED| class constructor takes the name
of a \ck{} file or a Python dictionary containing the contents of a \ck{} file
as its argument. When the file or dictionary is loaded, \pk{} validates its
format and contents using the Python package Cerberus \autocite{cerberus}. The
schema used for validation of the \ck{} files is available publicly in the
\texttt{schemas} directory of the PyKED source code
repository.\footnote{\url{https://github.com/pr-omethe-us/PyKED/tree/master/pyked/schemas}}
The schema comprises multiple YAML files to ease its extension to describe other
experimental measurements.

The fields of the \ck{} file are stored as instance attributes of the
\pybox|ChemKED| class. The following attributes are available:
\begin{itemize}
    \item \pybox|chemked_version|, \pybox|file_version|, \pybox|file_author|,
    \pybox|experiment_type|: Store the ``meta'' values from the \ck{} file
    \item \pybox|reference|: Instance of a \pybox|namedtuple| containing all
    the information from the literature reference associated with the data
    \item \pybox|apparatus|: Instance of a \pybox|namedtuple| containing all
    the information about the apparatus used to perform the experiment
    \item \pybox|datapoints|: A Python \pybox|list| of \pybox|DataPoint| instances
\end{itemize}

The \pybox|DataPoint| class stores the information associated with a single data
point in the \ck{} file (i.e., a single element of the \yabox|datapoints|
sequence). Similar to the \pybox|ChemKED| class,
the \pybox|DataPoint| stores information as instance attributes:
\begin{itemize}
    \item \pybox|equivalence_ratio|: The value of the equivalence ratio, if
    present. For informational purposes only---no validation is done of the
    value.
    \item \pybox|composition|: A list of dictionaries of the species and their
    respective amounts. The values are validated so that \yabox|mole percent|,
    \yabox|mole fraction|, or \yabox|mass fraction| cannot be mixed for a single
    data point, and so that the sum of the values is approximately 1.0, or
    100.0 for \yabox|mole percent|.
    \item \pybox|composition_type|: A string indicating the type of composition
    information for the data point---one of \pybox|'mole percent'|,
    \pybox|'mole fraction'|, or \pybox|'mass fraction'|
    \item \pybox|ignition_type|: A dictionary specifying the method of the
    measurement of ignition delay
    \item \pybox|volume_history|: If the \yabox|volume-history| of an RCM
    experiment is provided in the \ck{} file, it is stored in this attribute as
    a \pybox|namedtuple|, and the actual values are stored in NumPy arrays
    \autocite{vanderWalt:2011np}
\end{itemize}
Other instance attributes are stored as instances of the \pybox|Quantity| class from the
Pint \autocite{Grecco2016} package, possibly with an associated uncertainty. These include the
\pybox|ignition_delay|, \pybox|temperature|, \pybox|pressure|, \pybox|pressure_rise|,
\pybox|compression_time|, \pybox|compressed_temperature|, and \pybox|compressed_pressure|
attributes, each of which represents the similarly-named field in the \ck{} schema.

The \pybox|DataPoint| class defines two instance methods:
\pybox|get_cantera_mole_fraction()| and \pybox|get_cantera_mass_fraction()|.
These methods output the composition of the reactant mixture to a
format that can be used to set the composition of a
Cantera~\pybox|Solution|~\autocite{Cantera:2.3.0}.
The \pybox|composition| specification does not contain the molecular
weights of the components, so conversion between mole fractions and mass
fractions is not currently possible. (Future versions of \pk{} may support this
feature via, e.g., online lookup of species using their InChI\slash SMILES
identifiers.)

The \pybox|ChemKED| class defines three instance methods:
\pybox|get_dataframe()|, \pybox|write_file()|, and \pybox|convert_to_ReSpecTh()|.
The \pybox|get_dataframe()| method returns an instance of a
Pandas \pybox|DataFrame|~\autocite{pandas} that contains the information in the
list of \pybox|DataPoint|s. The user can specify the columns included in the
\pybox|DataFrame| by passing a list of column names to the
\pybox|output_columns| argument of the \pybox|get_dataframe()| method. The
possible columns are not case-sensitive and are:
\noindent\begin{itemize*}
    \item \pybox|'Temperature'|
    \item \pybox|'Pressure'|
    \item \pybox|'Ignition Delay'|
    \item \pybox|'Composition'|
    \item \pybox|'Equivalence Ratio'|
    \item \pybox|'Reference'|
    \item \pybox|'Apparatus'|
    \item \pybox|'Experiment Type'|
    \item \pybox|'File Author'|
    \item \pybox|'File Version'|
    \item \pybox|'ChemKED Version'|
\end{itemize*}

In addition, specific fields from the \pybox|`Reference'| and \pybox|`Apparatus'|
columns can be included by specifying the name after a colon. These options
are:
\noindent\begin{itemize*}
    \item \pybox|'Reference:Volume'|
    \item \pybox|'Reference:Journal'|
    \item \pybox|'Reference:DOI'|
    \item \pybox|'Reference:Authors'|
    \item \pybox|'Reference:Detail'|
    \item \pybox|'Reference:Year'|
    \item \pybox|'Reference:Pages'|
    \item \pybox|'Apparatus:Kind'|
    \item \pybox|'Apparatus:Facility'|
    \item \pybox|'Apparatus:Institution'|
\end{itemize*}

Only the first author is included in the \pybox|DataFrame| when \pybox|Reference| or
\pybox|Reference:Authors| is selected because the whole author list may be
quite long.

The \pybox|write_file()| method writes a new \ck{} YAML file based on the instance attributes
of the class, while \pybox|convert_to_ReSpecTh()| converts the \pybox|ChemKED|
instance to a ReSpecTh XML file. In the latter case, some information may be lost or the conversion
may not be possible, as \ck{} files support some data that ReSpecTh files do not support. We note
that the reverse is also true: not all ReSpecTh files may be
equivalently converted to \ck{} files because not
all of the data that can be stored in a ReSpecTh format is (currently) supported by the \ck{} format.

\pk{}~\autocite{PyKED} relies on well-established scientific Python software tools.
These include NumPy~\autocite{vanderWalt:2011np} and Pandas~\autocite{pandas,McKinney2010}
for array manipulation, Pint~\autocite{Grecco2016} for interpreting and converting
between units, PyYAML~\autocite{pyyaml} for parsing YAML
files, and Cerberus for validating \ck{} files~\autocite{cerberus}.
Furthermore, \pk{} includes extensive unit, integration, and functional tests
that ensure all aspects of the software operate as intended; others have pointed
out the importance of automated testing in scientific software~\autocite{Wilson:bestpractices}.
\pk{} relies on pytest~\autocite{pytest:3.0.1} for automated testing, and currently has
100\% line and branch coverage for all executable code in the repository. This
means that the tests execute every line of code at least once, and evaluate
every branch statement (e.g., if-then-else statements) to execute every condition
at least once.

Travis-CI\footnote{\url{https://travis-ci.org/pr-omethe-us/PyKED}} and
Appveyor\footnote{\url{https://ci.appveyor.com/project/Prometheus/pyked}} provide continuous
integration (CI) service for Linux, macOS, and Windows platforms. The CI services run the test suite
on every change to the source code, build and distribute binary installer packages for every
release, and build and publish the online documentation automatically. This automation ensures that
test failures are caught quickly, keeps the documentation up-to-date with the latest changes, and
streamlines the release of new versions.

\pk{} is licensed under the permissive, open-source BSD 3-clause license. The
source code is publicly available on GitHub at \url{https://github.com/pr-omethe-us/PyKED},
and versioned releases are automatically archived via Zenodo~\autocite{PyKED}.
\pk{} can be installed via binary packages or the source code; the online documentation
provides detailed installation instructions at
\url{https://pr-omethe-us.github.io/PyKED/install.html}.

%====================================================================
\section{Usage examples}
\label{sec:usage-example}

The following usage examples provide a guide to the use of \pk{}. They are by no means an exhaustive
treatment, and are meant to demonstrate the basic capabilities of the software. Both examples shown
below are also available as Jupyter Notebook files from the GitHub repository for
PyKED\footnote{\url{https://github.com/pr-omethe-us/PyKED/blob/master/docs/rcm-example.ipynb} and
\url{https://github.com/pr-omethe-us/PyKED/blob/master/docs/shock-tube-example.ipynb}} and in the
Supplementary Material associated with this article.

\subsection{RCM modeling with varying reactor volume}
\label{sec:rcm-modeling}

The \ck{} file that will be used in this example can be found in the
\texttt{tests} directory of the \pk{}
repository.\footnote{\url{https://github.com/pr-omethe-us/PyKED/blob/master/pyked/tests/testfile_rcm.yaml}}
Examining that file, we find the first section specifies the information about
the \ck{} file itself:
\begin{yamlbox}
file-author:
  name: Kyle E Niemeyer
  ORCID: 0000-0003-4425-7097
file-version: 0
chemked-version: 0.1.6
\end{yamlbox}
Then, we find the information regarding the article in the literature from which
this data was taken. In this case, the dataset comes from the work of
\textcite{Mittal2006a}:
%r
\begin{yamlbox}
reference:
  doi: 10.1002/kin.20180
  authors:
    - name: Gaurav Mittal
    - name: Chih-Jen Sung
      ORCID: 0000-0003-2046-8076
    - name: Richard A Yetter
  journal: International Journal of Chemical Kinetics
  year: 2006
  volume: 38
  pages: 516-529
  detail: Fig. 6, open circle
experiment-type: ignition delay
apparatus:
  kind: rapid compression machine
  institution: Case Western Reserve University
  facility: CWRU RCM
\end{yamlbox}
Finally, this file contains just a single datapoint, which describes the experimental
ignition delay, initial mixture composition, initial temperature, initial pressure,
compression time, ignition type, and volume history that specifies
how the volume of the reactor varies with time, for simulating the compression
stroke and post-compression processes:
\begin{yamlbox}
datapoints:
- temperature:
    - 297.4 kelvin
  ignition-delay:
    - 1.0 ms
  pressure:
    - 958.0 torr
  composition:
    kind: mole fraction
    species:
      - species-name: H2
        InChI: 1S/H2/h1H
        amount:
          - 0.12500
      - species-name: O2
        InChI: 1S/O2/c1-2
        amount:
          - 0.06250
      - species-name: N2
        InChI: 1S/N2/c1-2
        amount:
          - 0.18125
      - species-name: Ar
        InChI: 1S/Ar
        amount:
          - 0.63125
  ignition-type:
    target: pressure
    type: d/dt max
  compression-time:
    - 38.0 ms
  volume-history:
    time:
      units: s
      column: 0
    volume:
      units: cm3
      column: 1
    values:
      - [0.00E+000, 5.47669375000E+002]
      - [1.00E-003, 5.46608789894E+002]
      - [2.00E-003, 5.43427034574E+002]
      ...
\end{yamlbox}
The values for the \yabox|volume-history| are truncated here to save space. One application of the
data stored in this file is to perform a simulation using Cantera~\autocite{Cantera:2.3.0} to
calculate the ignition delay, including the facility-dependent effects represented in the volume
trace. All information required to perform this simulation is present in the \ck{} file, with the
exception of a chemical kinetic model for \ce{H2}\slash \ce{CO} combustion.

In Python, additional functionality can be imported into a script or session by the \pybox|import|
keyword. Cantera, NumPy, and PyKED must be imported into the session so that we can work with the
code. In the case of Cantera and NumPy, we will use many functions from these libraries, so we
assign them abbreviations (\pybox|ct| and \pybox|np|, respectively) for convenience. From PyKED, we
will only be using the \pybox|ChemKED| class, so this is all that is imported:
\begin{pythonbox}
import cantera as ct
import numpy as np
from pyked import ChemKED
\end{pythonbox}
Next, we have to load the ChemKED file and retrieve the first element of the \pybox|datapoints|
list. Although this file only encodes a single experiment, the \pybox|datapoints| attribute will
always be a list (in this case, of length 1). As mentioned previously, the elements of the
\pybox|datapoints| list are instances of the \pybox|DataPoint| class, which we store in the variable
\pybox|dp|.
\begin{pythonbox}
ck = ChemKED('testfile_rcm.yaml')
dp = ck.datapoints[0]
\end{pythonbox}
The initial temperature, pressure, and mixture composition can be read from the
instance of the \pybox|DataPoint| class. \pk{} uses instances of the Pint \pybox|Quantity| class to
store values with units, while Cantera expects a floating-point value in SI
units as input. Therefore, we use the built-in capabilities of Pint to convert
the units from those specified in the \ck{} file to SI units, and we use the \pybox|magnitude|
attribute of the \pybox|Quantity| class to take only the numerical part. We also retrieve the
initial mixture mole fractions in a format Cantera will understand:
\begin{pythonbox}
T_initial = dp.temperature.to('K').magnitude
P_initial = dp.pressure.to('Pa').magnitude
X_initial = dp.get_cantera_mole_fraction()
\end{pythonbox}

With these properties defined, we have to create the objects in Cantera that represent the physical
state of the system to be studied. In Cantera, the \pybox|Solution| class stores the thermodynamic,
kinetic, and transport data from the GRI Mech 3.0~\cite{GRI3.0} model included with Cantera.
After the \pybox|Solution| object is created, we can set the initial temperature, pressure,
and mole fractions using the \pybox|TPX| attribute of the \pybox|Solution| class:
\begin{pythonbox}
# Load the mechanism and set the initial state of the mixture
gas = ct.Solution('gri30.cti')
gas.TPX = T_initial, P_initial, X_initial
\end{pythonbox}

With the thermodynamic and kinetic data loaded and the initial conditions defined, we  need to
install the \pybox|Solution| instance into an \pybox|IdealGasReactor| which implements the equations
for mass, energy, and species conservation. In addition, we create a \pybox|Reservoir| to represent
the environment external to the reaction chamber. The input file used for the environment,
\verb|air.xml|, is included with Cantera and represents an average composition of air.
\begin{pythonbox}
# Create the reactor and the outside environment
reac = ct.IdealGasReactor(gas)
env = ct.Reservoir(ct.Solution('air.xml'))
\end{pythonbox}
To apply the effect of the volume trace to the \pybox|IdealGasReactor|, a \pybox|Wall| must be
installed between the reactor and environment and assigned a velocity. The \pybox|Wall| allows the
environment to do work on the reactor (or vice versa) and change the reactor's thermodynamic state;
we use a \pybox|Reservoir| for the environment because in Cantera, \pybox|Reservoir|s always have a
constant thermodynamic state and composition. Using a \pybox|Reservoir| accelerates the solution
compared to using two \pybox|IdealGasReactor|s, since the composition and state of the environment
are typically not necessary for the solution of autoignition problems. Although we do not show the
details here, a reference implementation of a class that computes the wall velocity given the volume
history of the reactor is available in CanSen~\autocite{cansen}, in the
\pybox|cansen.profiles.VolumeProfile| class.
\begin{pythonbox}
from cansen.profiles import VolumeProfile
# Retrieve the time and volume from the history in the datapoint
exp_time = dp.volume_history.time.magnitude
exp_volume = dp.volume_history.volume.magnitude
keywords = {'vproTime': exp_time, 'vproVol': exp_volume}
# Install the Wall between the reactor and environment
ct.Wall(reac, env, velocity=VolumeProfile(keywords))
\end{pythonbox}

Then, the \pybox|IdealGasReactor| is installed in a \pybox|ReactorNet|. The \pybox|ReactorNet|
implements the connection to the numerical solver (CVODES \autocite{hindmarsh_sundials:_2005} is
used in Cantera) to solve the energy and species equations. For this example, it is best practice
to set the maximum time step allowed in the solution to be the minimum time difference in
the time array from the volume trace:
\begin{pythonbox}
netw = ct.ReactorNet([reac])
netw.set_max_time_step(np.min(np.diff(exp_time)))
\end{pythonbox}

To calculate the ignition delay, we will follow the definition specified in the \ck{} file for
this experiment, where the experimentalists used the maximum of the time derivative of the pressure
to define the ignition delay. To calculate this derivative, we need to store the state variables
and the composition on each time step, so we initialize several Python lists to act as storage:
\begin{pythonbox}
# Initialize lists to store solution information
time = []
temperature = []
pressure = []
volume = []
mass_fractions = []
\end{pythonbox}

Finally, the problem is integrated using the \pybox|step| method of the \pybox|ReactorNet|. The
\pybox|step| method takes one timestep forward on each call, with step size determined by the CVODES
solver (CVODES uses an adaptive time-stepping algorithm). On each step, we add the relevant
variables to their respective lists. The problem is integrated until a user-specified end time, in
this case \SI{50}{\milli\second}, although in principle, the user could end the simulation on any
condition they choose:
\begin{pythonbox}
# Integrate for 50 ms
while netw.time < 0.05:
    time.append(netw.time)
    temperature.append(reac.T)
    pressure.append(reac.thermo.P)
    volume.append(reac.volume)
    mass_fractions.append(reac.Y)
    netw.step()
\end{pythonbox}

\begin{figure}[htbp]
    \centering
    \begin{subfigure}{0.7\textwidth}
        \centering
        \includegraphics[width=\linewidth]{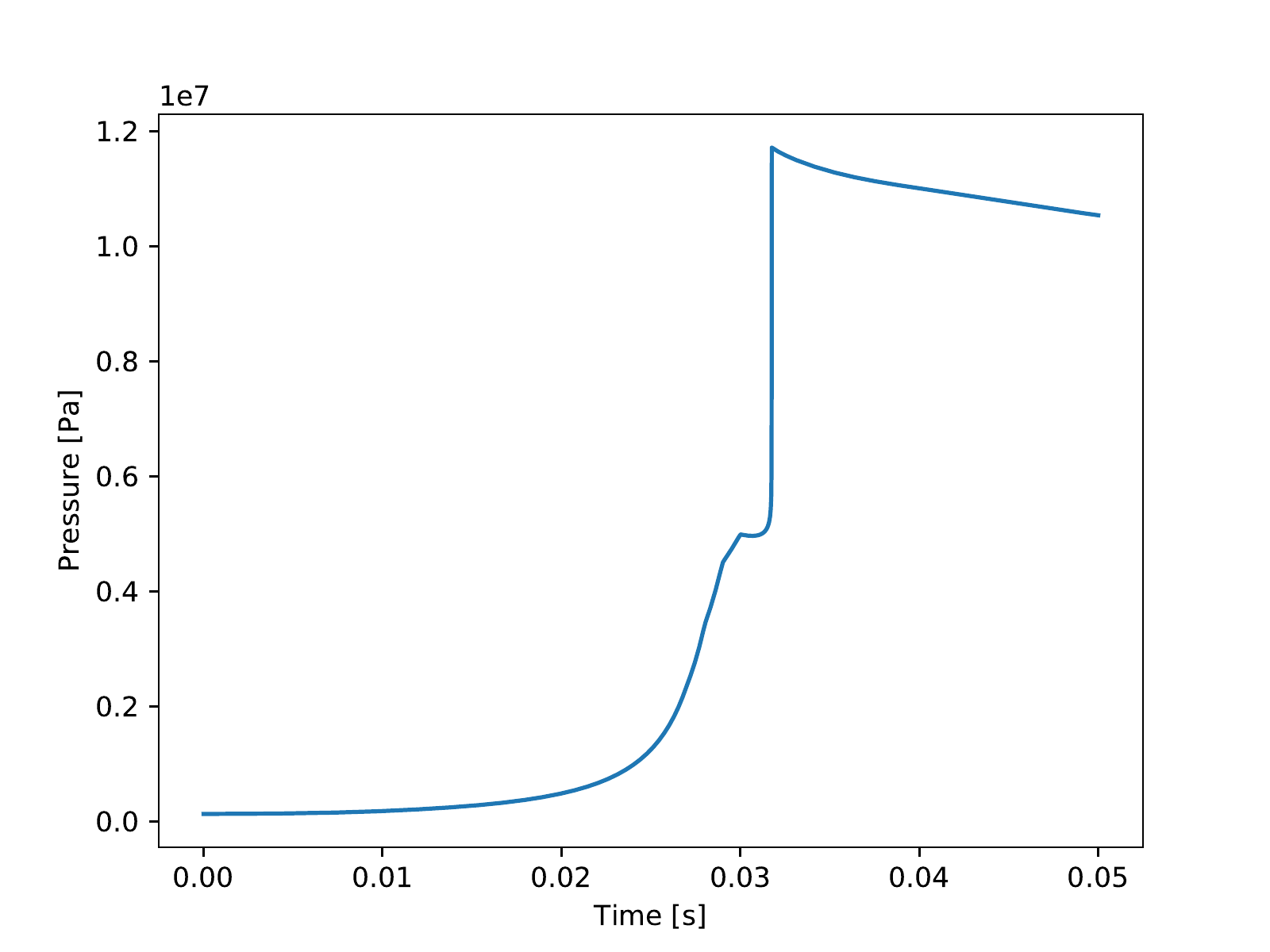}
        \caption{Simulated pressure trace}
    \end{subfigure}
    \\
    \begin{subfigure}{0.7\textwidth}
        \centering
        \includegraphics[width=\linewidth]{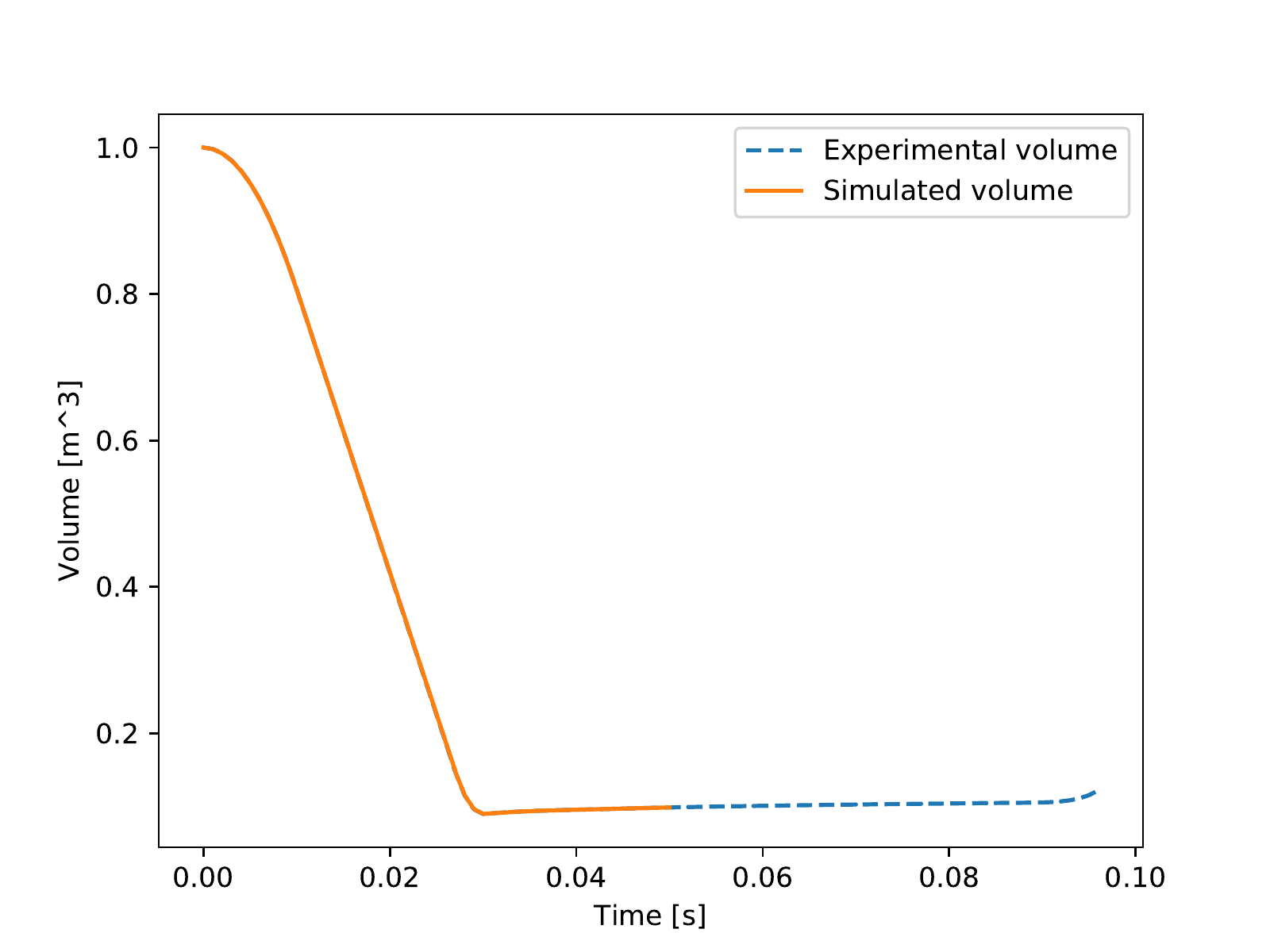}
        \caption{Experimental and simulated volume histories}
    \end{subfigure}
    \caption{Results from RCM example}
    \label{fig:rcm}
\end{figure}

At this point, the user would post-process the information in the \pybox|pressure| list to calculate
the derivative by whatever algorithm they choose. Here, we plot pressure versus time from the
simulation, and compare the simulated and experimental volume traces,
using the Matplotlib library~\cite{matplotlib} and shown in Figure~\ref{fig:rcm}:
\begin{pythonbox}
import matplotlib.pyplot as plt

plt.figure()
plt.plot(time, pressure)
plt.ylabel('Pressure [Pa]')
plt.xlabel('Time [s]')

plt.figure()
plt.plot(exp_time, exp_volume/exp_volume[0], label='Experimental volume', linestyle='--')
plt.plot(time, volume, label='Simulated volume')
plt.legend(loc='best')
plt.ylabel('Volume [m^3]')
plt.xlabel('Time [s]')
\end{pythonbox}

\subsection{Shock tube modeling with constant volume}
\label{sec:shock-tube}

The \ck{} file used in this example can be found in the \texttt{tests} directory of the \pk{}
repository.\footnote{\url{https://github.com/pr-omethe-us/PyKED/blob/master/pyked/tests/testfile_st_p5.yaml}}
The data in this file comes from Stranic et al.~\autocite{Stranic:2012}, describing
shock-tube ignition delays for \textit{tert}-butanol. We have omitted the file
meta information below for space; the format is largely similar to the example in
Section~\ref{sec:rcm-modeling}.
This \ck{} file specifies multiple data points with some common
conditions, including a common mixture composition and common definition of
ignition delay. Therefore, a \yabox|common-properties| section is specified, followed by the
\yabox|datapoints| list (as before, we have truncated the \yabox|datapoints| list for space):
\begin{yamlbox}
common-properties:
  composition: &comp
    kind: mole fraction
    species:
      - species-name: t-butanol
        InChI: 1S/C4H10O/c1-4(2,3)5/h5H,1-3H3
        amount:
          - 0.003333333
      - species-name: O2
        InChI:  1S/O2/c1-2
        amount:
          - 0.04
      - species-name: Ar
        InChI:  1S/Ar
        amount:
          - 0.956666667
  ignition-type: &ign
    target: OH*
    type: 1/2 max
datapoints:
  - temperature:
      - 1459 kelvin
    ignition-delay:
      - 347 us
    pressure:
      - 1.60 atm
    composition: *comp
    ignition-type: *ign
    equivalence-ratio: 0.5
  - temperature:
      - 1389 kelvin
    ignition-delay:
      - 756 us
    pressure:
      - 1.67 atm
    composition: *comp
    ignition-type: *ign
    equivalence-ratio: 0.5
    ...
\end{yamlbox}

In this example, we will run constant-volume simulations at each
pressure and temperature condition in the \yabox|datapoints| list. Once again,
the \ck{} file specifies all information required for the simulations except
for the chemical kinetic model, and Cantera can be used to simulate autoignition. The setup steps
match those from the previous example, with one exception: in this example, we also import
Python's built-in multiprocessing library so that we can run the simulations on multiple cores.
We import the \pybox|Pool| class, which offers tools to manage a set of jobs on a
pool of processors:
\begin{pythonbox}
import cantera as ct
from multiprocessing import Pool
from pyked import ChemKED
\end{pythonbox}

Then, we define a function that will be mapped onto each job. This function takes the initial
temperature, pressure, and mole fractions as input and returns the simulated ignition delay time,
defined as the time the temperature increases by \SI{400}{\K} over the initial temperature, a
simplified definition for this example. In general, the user could process the mole fraction of
\ce{OH^*} (provided that the kinetic model includes accurate chemistry for \ce{OH^*}) to match the
definition of ignition delay in the experiments. We use the chemical kinetic model
for butanol isomers from Sarathy et al.~\cite{Sarathy:2012fj}, represented as
\verb|LLNL_sarathy_butanol.cti|:
\begin{pythonbox}
def run_simulation(T, P, X):
    gas = ct.Solution('LLNL_sarathy_butanol.cti')
    gas.TPX = T, P, X
    reac = ct.IdealGasReactor(gas)
    netw = ct.ReactorNet([reac])
    while reac.T < T + 400:
        netw.step()

    return netw.time
\end{pythonbox}

Then, we load the ChemKED file and generate a list of initial conditions that will be mapped onto
the \pybox|run_simulation()| function. We first define a convenience function to collect the input
from a single datapoint and return a Python tuple with the conditions (tuples are lightweight groups
of data in Python). Then, we use the built-in \pybox|map| function to apply the
\pybox|collect_input| function to each of the elements in the \pybox|ck.datapoints| list:
\begin{pythonbox}
ck = ChemKED('Stranic2012-tbuoh.yaml')

def collect_input(dp):
    T_initial = dp.temperature.to('K').magnitude
    P_initial = dp.pressure.to('Pa').magnitude
    X_initial = dp.get_cantera_mole_fraction()
    return (T_initial, P_initial, X_initial)

initial_conditions = list(map(collect_input, ck.datapoints))
\end{pythonbox}

\begin{figure}[htbp]
    \centering
    \includegraphics[width=0.7\textwidth]{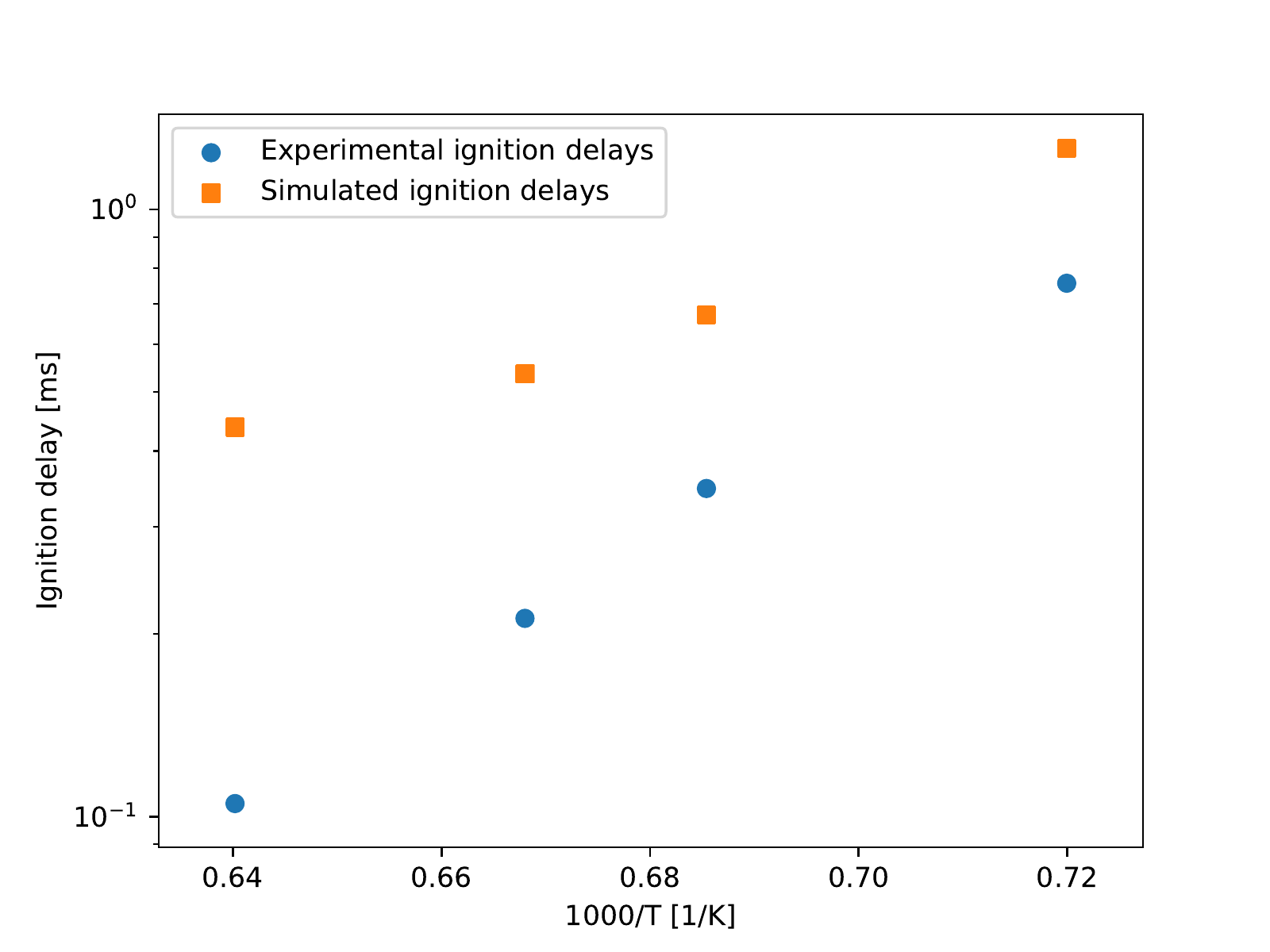}
    \caption{Comparison of experimental and simulated ignition delays of
    \textit{tert}-butanol from shock tube example.}
    \label{fig:ignition-delays}
\end{figure}

Finally, we create the processor \pybox|Pool| (with four processes) and send the jobs out to run:
\begin{pythonbox}
with Pool(processes=4) as pool:
    ignition_delays = pool.starmap(run_simulation, initial_conditions)

for (T, P, X), tau in zip(initial_conditions, ignition_delays):
    print(f'The ignition delay for T_initial={T} K, P_initial={P} Pa is: {tau} seconds')
\end{pythonbox}
The simulated ignition delay results are returned in the \pybox|ignition_delays| list. The results
are printed to the screen using a Python formatted string (f-string). We can also visually
compare the results using Matplotlib~\cite{matplotlib}, as
Figure~\ref{fig:ignition-delays} shows:
\begin{pythonbox}
import matplotlib.pyplot as plt

inv_temps = [1000/i[0] for i in initial_conditions]
exp_ignition_delays = [dp.ignition_delay.to('ms').magnitude for dp in ck.datapoints]
sim_ignition_delays = np.array(ignition_delays)*1.0E3

plt.figure()
plt.scatter(inv_temps, exp_ignition_delays, label='Experimental ignition delays')
plt.scatter(inv_temps, sim_ignition_delays, label='Simulated ignition delays', marker='s')
plt.legend(loc='best')
plt.yscale('log')
plt.ylabel('Ignition delay [ms]')
plt.xlabel('1000/T [1/K]')
\end{pythonbox}

%====================================================================
\section{Conclusions and future work}

In this article, we presented the \ck{} data format for describing measurements
taken from fundamental combustion experiments, recognizing that the community
has a need for an open and standardized data serialization format. \ck{} files
are formatted using the YAML language and are plain-text, human- and
machine-readable, and easy to construct.

We also presented a Python-based tool, \pk{}, for validating and working
with \ck{} files. \pk{} provides the reference implementation of the validator
for \ck{} files and utilizes several common packages from Python's scientific
computing community. \pk{} and \ck{} currently support ignition delay
measurements from rapid compression machines and shock tubes, including
facility-specific effects from each type of experiment.

Finally, we presented several examples using \pk{} to interpret and interact with data
stored in \ck{} files. Both examples are also available online as Jupyter Notebook files in the
GitHub repository and in the Supplementary Material. The first example simulates a
rapid compression machine experiment, including the facility-specific effects critical to
accurately predict ignition delay experiments. The second example demonstrates the use of built-in
Python libraries to automatically simulate several shock tube experiments in parallel,
with initial conditions loaded directly from the \ck{} file.

We are actively developing \ck{} and \pk{}, and we welcome contributions from the community.
All development occurs under the BSD 3-clause open source license and the code is housed
on GitHub at \url{https://github.com/pr-omethe-us/PyKED}. Future directions for development are
outlined in the public roadmap.\footnote{\url{https://github.com/pr-omethe-us/PyKED/wiki/Roadmap}}
The highest priority issues are currently adding support for other types of fundamental experiments,
including speciation measurements in shock tubes, jet-stirred reactors, and flow reactors, and
laminar flame speed and flame extinction measurements. Questions, comments, or suggestions are
welcomed and can be posted as issues in the GitHub repository or emailed to the authors. In
particular, we welcome proposals for new experimental types; see issue \#60 in the \pk{} repository
for an example of how this may be
done.\footnote{\url{https://github.com/pr-omethe-us/PyKED/issues/60}}

\printbibliography

\end{document}